%% file: ADiC_MILCOM19.tex
\documentclass[10pt,conference,letterpaper]{IEEEtran}
\usepackage[USenglish]{babel}
\usepackage{times,amsmath}
\usepackage[pdftex]{graphicx}
   \graphicspath{{./pdf/}{./png/}{./jpeg/}{./eps/}{./png_old/}}
   \DeclareGraphicsExtensions{.pdf,.png,.jpeg,.eps}
\usepackage{cite}
\usepackage{url}
\usepackage{fmtcount}
\usepackage{color}
\usepackage[tight,footnotesize]{subfigure}
\usepackage{multicol}
\usepackage{amssymb}
\usepackage{wasysym}
\usepackage{dblfloatfix} 
\usepackage{calc}
\usepackage{fix-cm}
\usepackage[activate={true,nocompatibility},final,tracking=true,factor=500,stretch=40,shrink=40]{microtype}
\def\cal{\fam2}
\renewcommand{\d}{{\mathrm d}}

\newcommand{\sgn}{{\mathrm{sgn}}}

\newcommand{\half}{{\frac{1}{2}}}

\newcommand{\BalphaPM}{{{\cal{B}}^{\boldmath \alpha_+}_{\boldmath \alpha_-}}}
\definecolor{DarkRed}{rgb}{0.5,0,0}
\definecolor{DarkGreen}{rgb}{0,0.5,0}
\definecolor{DarkerGreen}{rgb}{0,0.3333,0}
\definecolor{DarkBlue}{rgb}{0,0,0.75}
\definecolor{RoyalBlue}{rgb}{0,0.1373,0.4000}
\definecolor{NavyBlue}{rgb}{0,0,0.5020}
\definecolor{CobaltBlue}{rgb}{0,0.2784,0.6706}
\definecolor{lightlightgray}{rgb}{0.96875,0.96875,0.96875}
\definecolor{cyan}{rgb}{0,1,1}
\newcommand{\beginlabel}[2]{%
\begin{#1}\label{#2}}
\makeatletter
\def\ps@IEEEtitlepagestyle{%
\def\@oddhead{\mbox{}\scriptsize\rightmark Nikitin and Davidchack \hfil Complementary Intermittently Nonlinear Filtering for Mitigation of Hidden Outlier Interference \hfil \thepage}%
\def\@evenhead{\scriptsize\thepage \hfil Nikitin and Davidchack \leftmark\mbox{}}%
\def\@oddfoot{\scriptsize MILCOM~2019 draft \hfil \today \hfil \includegraphics[width=20mm]{NonLinear_logo_RoyalBlue_pst.png} \hspace*{2ex} \includegraphics[width=20mm]{UoL-Logo-Mono.png}}%
\def\@oddfoot{\scriptsize MILCOM~2019 draft \hfil \today \hfil \includegraphics[width=20mm]{NonLinear_logo_RoyalBlue_pst.png} \hspace*{2ex} \includegraphics[width=20mm]{UoL-Logo-Mono.png}}}
\makeatother
\makeatletter
\def\ps@headings{%
\def\@oddhead{\mbox{}\scriptsize\rightmark Nikitin and Davidchack \hfil Complementary Intermittently Nonlinear Filtering for Mitigation of Hidden Outlier Interference \hfil \thepage}%
\def\@evenhead{\scriptsize\thepage \hfil Nikitin and Davidchack \leftmark\mbox{}}%
\def\@oddfoot{\scriptsize MILCOM~2019 draft \hfil \today \hfil \includegraphics[width=20mm]{NonLinear_logo_RoyalBlue_pst.png} \hspace*{2ex} \includegraphics[width=20mm]{UoL-Logo-Mono.png}}%
\def\@oddfoot{\scriptsize MILCOM~2019 draft \hfil \today \hfil \includegraphics[width=20mm]{NonLinear_logo_RoyalBlue_pst.png} \hspace*{2ex} \includegraphics[width=20mm]{UoL-Logo-Mono.png}}}
\makeatother
\begin{document}
\pagestyle{plain}
\title{Complementary Intermittently Nonlinear Filtering for Mitigation of Hidden Outlier Interference}
\author{\IEEEauthorblockN{Alexei V. Nikitin}
\IEEEauthorblockA{
Nonlinear LLC\\
Wamego, Kansas, USA\\
E-mail: avn@nonlinearcorp.com}
\and
\IEEEauthorblockN{Ruslan L. Davidchack}
\IEEEauthorblockA{Dept. of Mathematics, U. of Leicester\\
Leicester, UK\\
E-mail: rld8@leicester.ac.uk}}
\maketitle
\thispagestyle{plain}
\begin{abstract}
When interference affecting various communication and sensor systems contains clearly identifiable outliers (e.g. an impulsive component), it can be efficiently mitigated in real time by intermittently nonlinear filters developed in our earlier work, achieving improvements in the signal quality otherwise unattainable. However, apparent amplitude outliers in the interference can disappear and reappear due to various filtering effects, including fading and multipass, as the signal propagates through media and/or the signal processing chain. In addition, the outlier structure of the interference can be obscured by strong non-outlier interfering signals, such as thermal noise and/or adjacent channel interference, or by the signal of interest itself. In this paper, we first outline the overall approach to using intermittently nonlinear filters for in-band, real-time mitigation of such interference with hidden outlier components in practical complex interference scenarios. We then introduce Complementary Intermittently Nonlinear Filters (CINFs) and focus on the particular task of mitigating the outlier noise obscured by the signal of interest itself. We describe practical implementations of such nonlinear filtering arrangements for mitigation of hidden outlier interference, in the process of analog-to-digital conversion, for wide ranges of interference powers and the rates of outlier generating events. To emphasize the effectiveness and versatility of this approach, in our examples we use particularly challenging waveforms that severely obscure low-amplitude outlier noise, such as broadband chirp signals (e.g. used in radar, sonar, and spread-spectrum communications) and ``bursty," high crest factor signals (e.g. OFDM).
\end{abstract}
\maketitle
\section{Methodology for real-time mitigation of hidden outlier interference} \label{sec:methodology}
At any given frequency, a linear filter affects all signals proportionally. Thus, when linear filtering is used to suppress interference, the resulting signal quality is largely invariant to a particular makeup of the interfering signal and depends mainly on the total power and the spectral composition of the interference in the passband of interest. On the other hand, properly implemented intermittently nonlinear filtering enables in-band, real-time reduction of interference with distinct outlier components, achieving mitigation levels unattainable by linear filters~\cite{Nikitin19hidden, Nikitin18ADiC-ICC, Nikitin19ADiCpatent}. While being nonlinear in general, intermittently nonlinear filters largely behave linearly. They exhibit nonlinear behavior only intermittently, in response to noise outliers, thus avoiding the detrimental effects, such as instabilities and intermodulation distortions, often associated with nonlinear filtering.

\begin{figure}[!b]
\centering{\includegraphics[width=8.6cm]{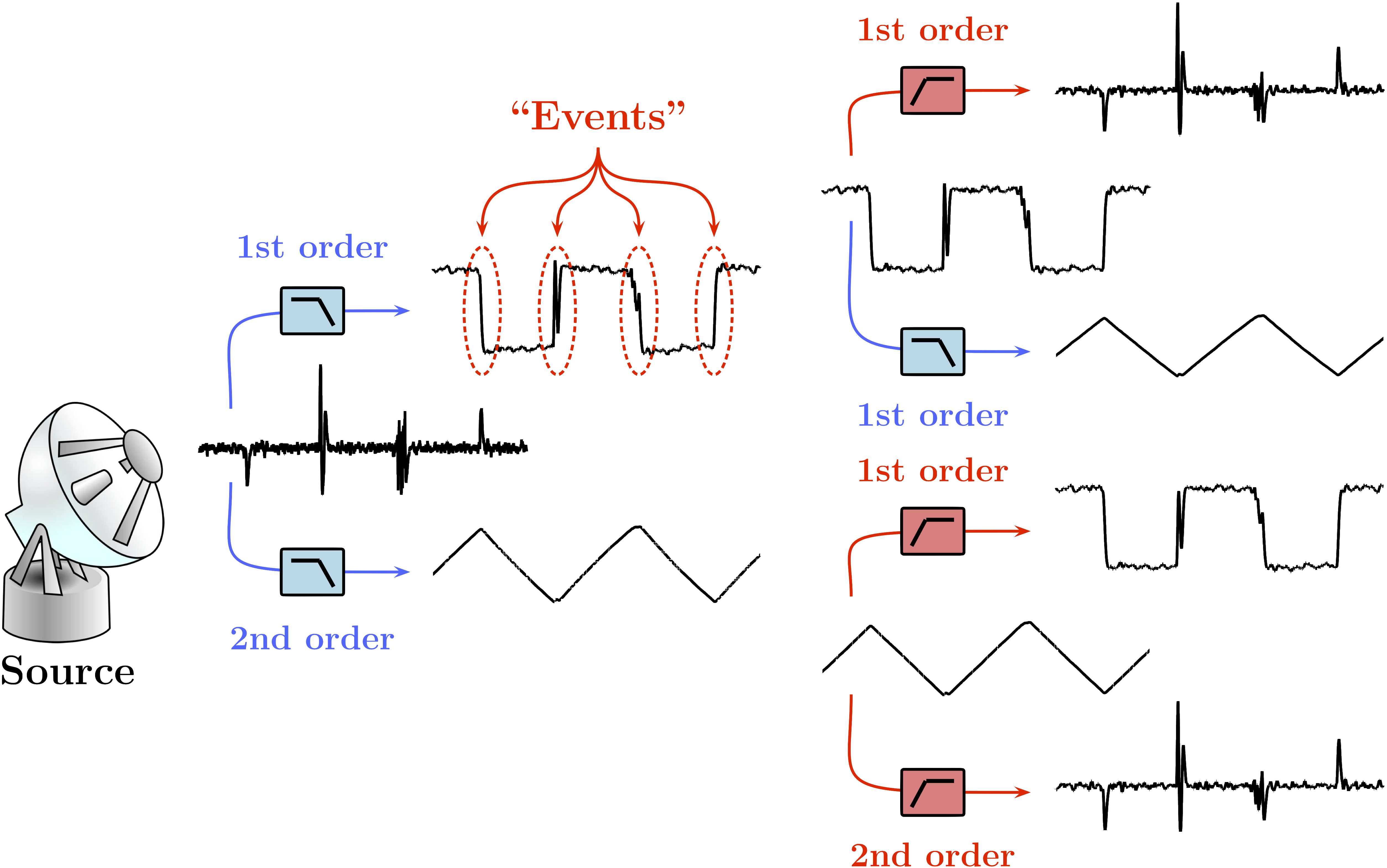}}
\caption{Outlier noise produced by ``events" separated by ``inactivity" when observed at wide bandwidth.
\label{fig:morphing}}
\end{figure}

\begin{figure}[!t]
\centering{\includegraphics[width=8.6cm]{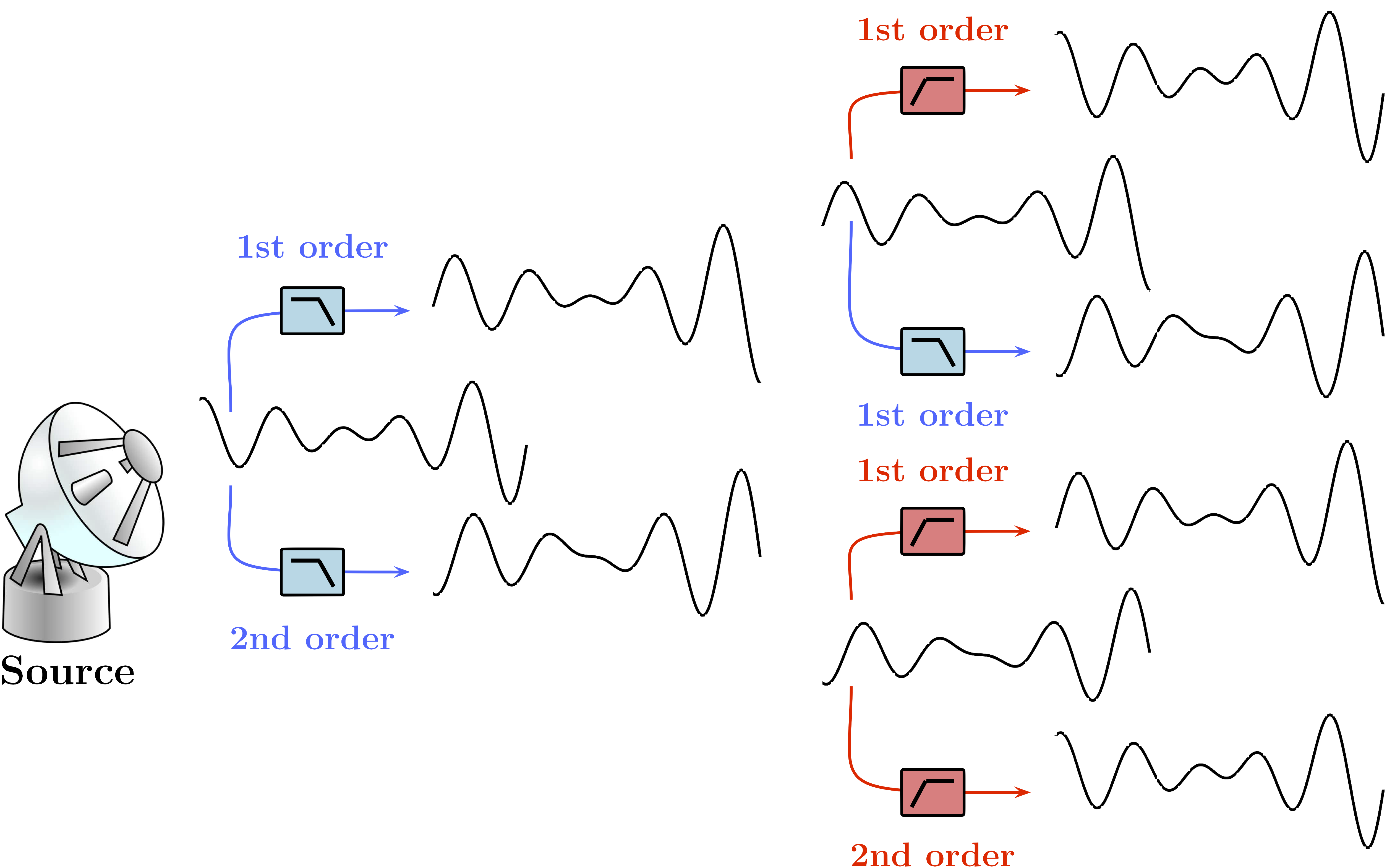}}
\caption{Same linear filtering does not produce outliers in band-limited signal of interest.
\label{fig:bandlimited}}
\end{figure}

Many interfering signals originating from natural and technogenic (man-made) phenomena can contain components that are produced by some ``countable" or ``discrete," relatively short duration events separated by relatively longer periods of inactivity. Given the same sequence of events, the time-domain appearance of such components can vary greatly, depending on the coupling mechanisms and the system's and propagation media's filtering properties. For example, while all three broadband signals shown in Fig.~\ref{fig:morphing} are produced by the same sequence of events, their time domain appearances, as well as the spectral densities, are very different due to different system responses and/or filtering effects, and only one of these signals contains clearly visible time-domain outliers. Nevertheless, these signals represent the same source and they can be morphed into each other by simple 1st or 2nd order filters. On the other hand, a signal of interest would be typically constructed to be band-limited (if for no other reason than to reduce its out-of-band interference with other signals), and confined to a band narrower than that available for observation of interference. As illustrated in Fig.~\ref{fig:bandlimited}, the same filtering applied to the signal of interest, while changing the appearance of the signal, is less likely to produce distinct outliers.

\begin{figure}[!b]
\centering{\includegraphics[width=8.6cm]{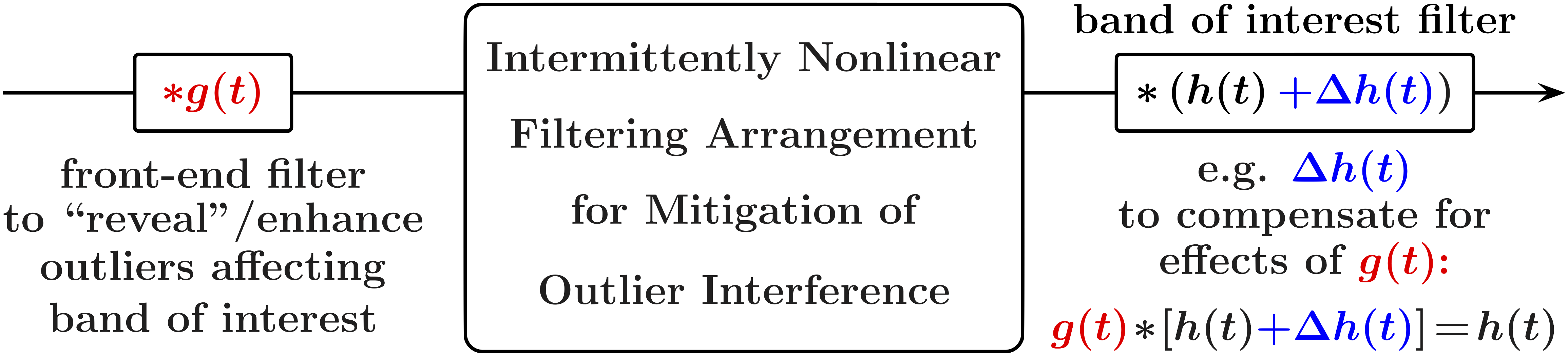}}
\caption{{\bf Addressing complex interference scenarios:~}%
Front-end linear filter reveals/enhances outliers affecting band of interest.
\label{fig:complex}}
\end{figure}

Hence, when observation bandwidth sufficiently larger than that of the signal of interest is available, various combinations of linear filters can be used to increase the difference between the temporal and/or amplitude structures of the interference and the signal of interest, enhancing the outlier components of the interference and enabling its mitigation by intermittently nonlinear filtering. Such combinations of front-end linear filters can be used not only for ``revealing" amplitude outliers hidden in broadband interference, but also for suppressing strong non-outlier interfering signals that may otherwise obscure these outliers, e.g. adjacent channel interference. This approach is summarized in Fig.~\ref{fig:complex}, where a linear filter is employed ahead of the intermittently nonlinear filter to reveal and/or enhance the outliers affecting the band of interest. Subsequently, if needed, the ``band of interest" filter (e.g. the digital decimation filter) can be modified to compensate for the impact of the front-end filter on the signal of interest. Several examples illustrating this technique can be found in~\cite{Nikitin19hidden, Nikitin19ADiCpatentCIPs}.


\subsection{Complementary intermittently nonlinear filtering} \label{subsec:CINF}
\begin{figure}[!t]
\centering{\includegraphics[width=8.6cm]{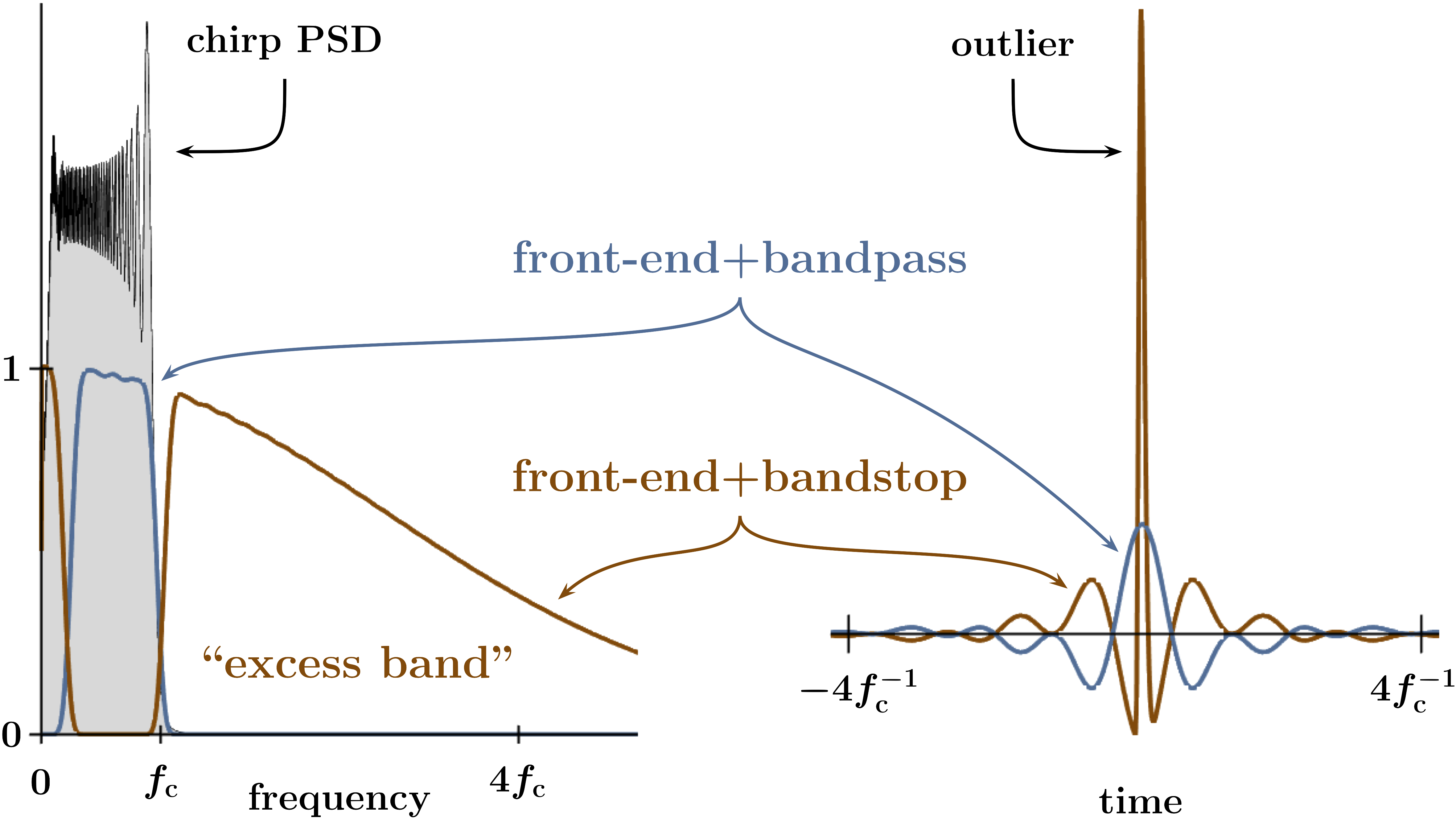}}
\caption{Impulse and frequency responses of linear filters used in Fig.~\ref{fig:chirp CAF}.
\label{fig:excess band}}
\end{figure}

\begin{figure}[!b]
\centering{\includegraphics[width=8.6cm]{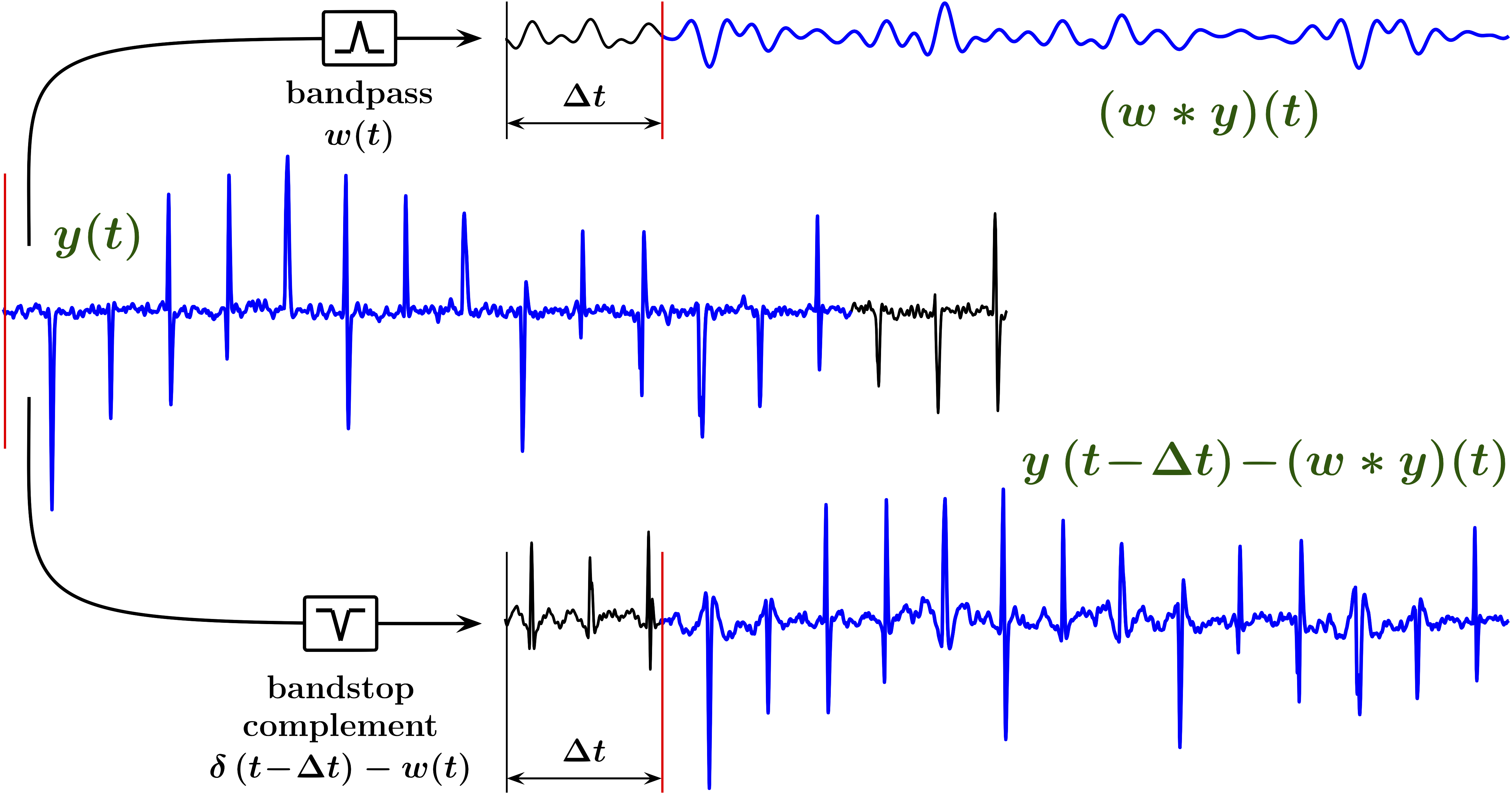}}
\caption{Noise filtered with bandpass and bandstop filters used in Fig.~\ref{fig:chirp CAF}.
\label{fig:filtered noise}}
\end{figure}

\begin{figure*}[!t]
\centering{\includegraphics[width=16.4cm]{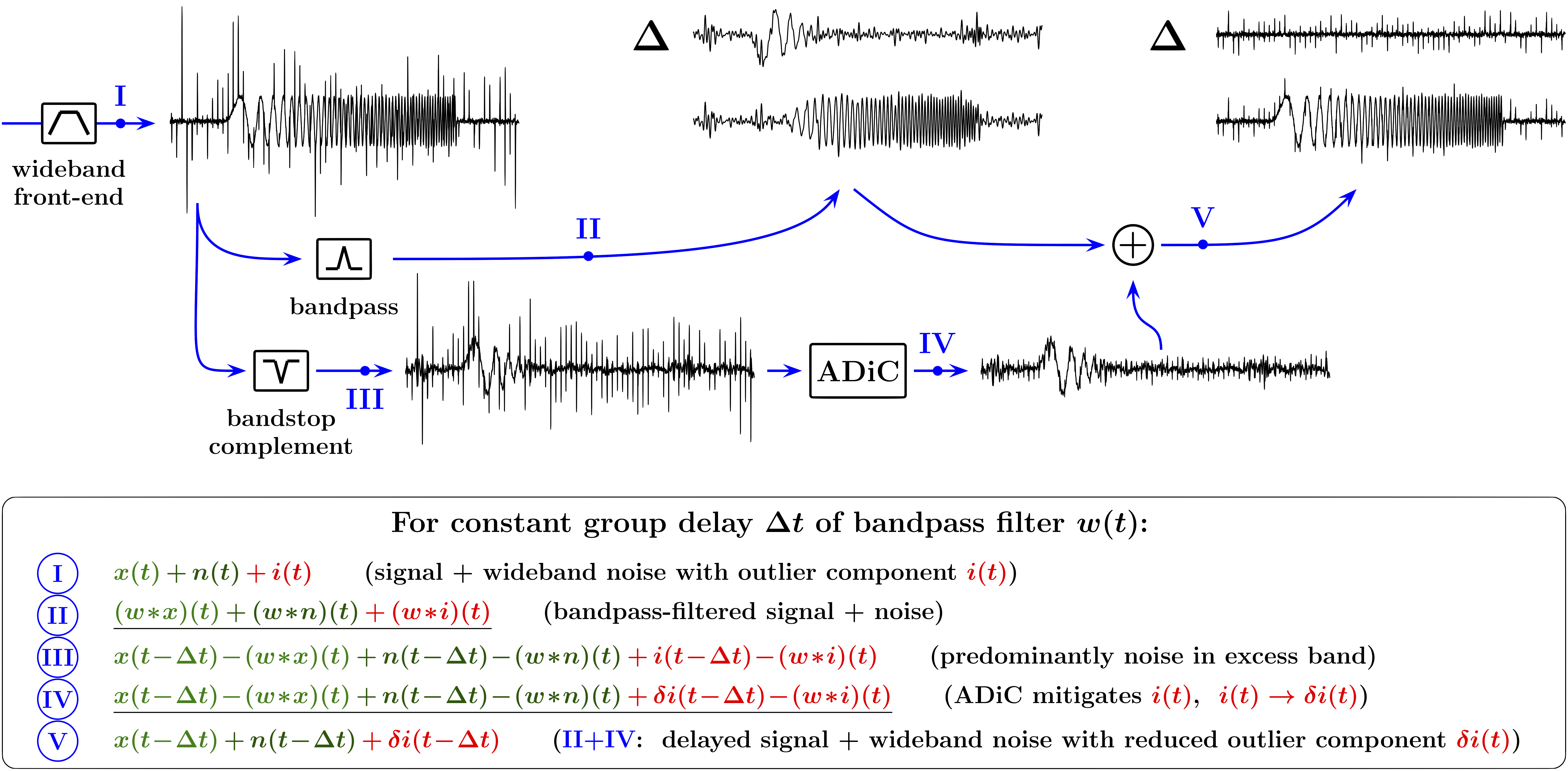}}
\caption{Complementary ADiC filtering (CAF) for removing wideband noise outliers while preserving band-limited signal of interest.
\label{fig:chirp CAF}}
\end{figure*}

\begin{figure*}[!t]
\centering{\includegraphics[width=12.4cm]{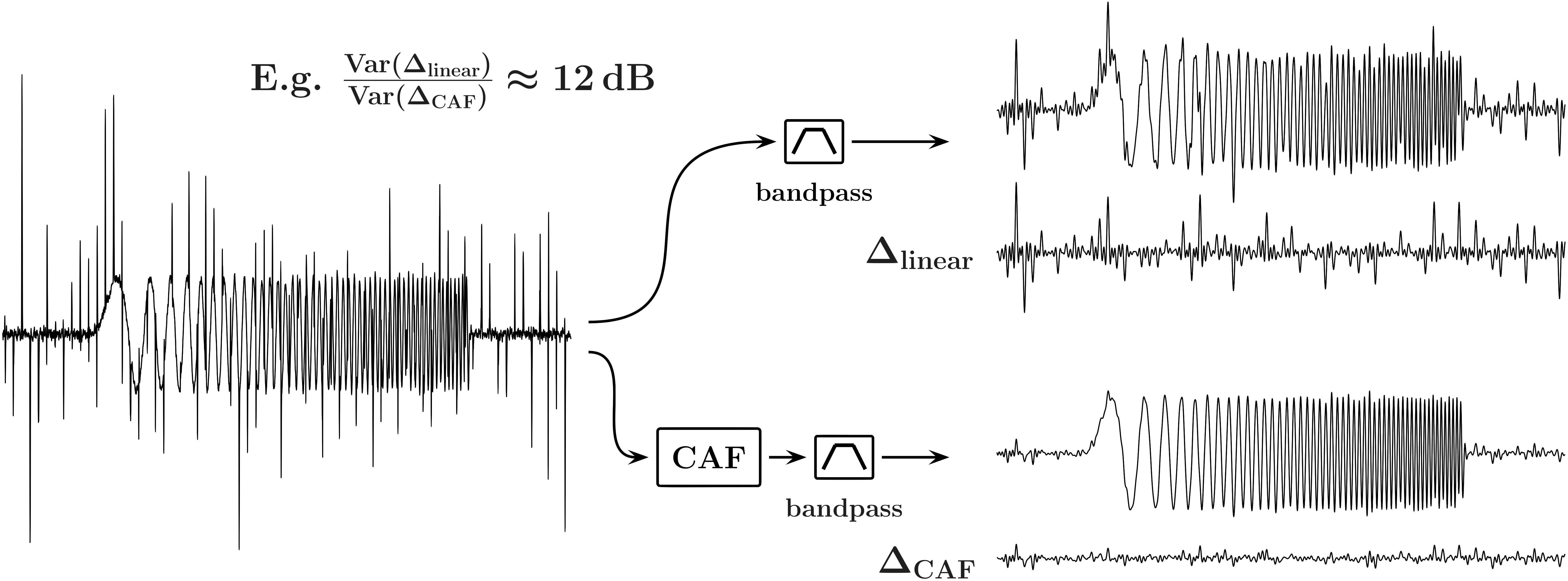}}
\caption{Comparison between outcomes of linear and CAF-based outlier noise filtering.
\label{fig:chirp example}}
\end{figure*}

The outlier interference can be further obscured by the signal of interest itself, especially when the typical signal amplitude is comparable with, or larger than the typical amplitude of the interference outliers. Examples of particularly challenging waveforms that severely obscure low-amplitude outlier noise include broadband chirp signals used in radar, sonar, and spread-spectrum communications, and ``bursty," high crest factor signals such as those used in OFDM systems. This challenge can be addressed by Complementary Intermittently Nonlinear Filtering (CINF) introduced in~\cite{Nikitin19hidden, Nikitin19ADiCpatentCIPs}. This approach capitalizes on the ``excess band" observation of wideband outlier noise for its efficient in-band mitigation. This significantly extends the mitigation range, in terms of both the rates of the outlier generating events and the mitigable signal-to-noise ratios (SNRs), in comparison with the mitigation techniques focused on the apparent in-band effects of outlier interference. Here, in Figs.~\ref{fig:excess band} through~\ref{fig:chirp example} we provide a detailed illustration of such a complementary filtering arrangement for mitigation of outlier noise affecting a broadband chirp signal.


First, Fig.~\ref{fig:excess band} illustrates the impulse and frequency responses of the linear filters used in the example of Fig.~\ref{fig:chirp CAF}. The bandwidth of the front-end filter should be large relative to the bandwidth of the signal of interest, so that a sufficiently wide ``excess band" is available, and its time-bandwidth product should be sufficiently small so that the combined impulse response of the front-end filter cascaded with the bandstop filter contains a distinct outlier. The amount of excess bandwidth that can be allocated for outlier noise mitigation depends on the particular requirements and constraints placed on a system, and the excess bandwidth availability affects both the ``mitigable rates" (e.g. in terms of the rates of outlier generating events) and ``mitigable SNRs" (e.g. in terms of outlier-to-thermal noise powers) of the outlier interference. For example, for a random pulse train, when the ratio of the bandwidth and the pulse arrival rate becomes significantly smaller than the time-bandwidth product of a filter, the resulting signal becomes effectively Gaussian due to the so-called ``pileup effect"~\cite[e.g.]{Nikitin98ppileup}, making the impulsive noise completely disappear. The time-bandwidth product of a lowpass Bessel filter is approximately that of a Gaussian filter, ${2\log_2(2)/\pi}$, and thus~$\lambda_{\rm c}\approx 2.27B_0$ is the ``pileup threshold" rate of the front-end Bessel filter with the 3\,dB bandwidth~$B_0$. For outlier arrival rates significantly above~$\lambda_{\rm c}$ the outlier noise can no longer be efficiently mitigated.

Due to high slew rates, the higher-frequency portion of the chirp signal is the most effective in obscuring low-amplitude broadband outliers. Therefore the bandstop complement of the bandpass filter should significantly reduce these high frequencies, in order for the outlier interference affecting the high-frequency portion of the chirp to become more conspicuous. On the other hand, the stopband of the bandstop filter should remain sufficiently narrow, so that the outlier in the combined impulse response of the front-end filter cascaded with the bandstop filter remains distinct. Thus, in the examples of Figs.~\ref{fig:excess band} through~\ref{fig:chirp example}, the high-frequency edge~$f_{\rm c}$ of the stopband is chosen at approximately the highest frequency of the chirp signal, and the low-frequency edge is placed at approximately~$f_{\rm c}/5$. While such a bandstop filter reduces the average slew rate of a linear chirp by about an order of magnitude, its stopband remains relatively narrow in comparison with the passband of the front-end filter, and it will mainly preserve the outlier structure of the noise. This is illustrated in Fig.~\ref{fig:filtered noise}, where $w(t)$ is the impulse response of the bandpass filter with a constant group delay~$\Delta{t}$, and~$\delta(t)$ is the Dirac $\delta$-function~\cite{Dirac58principles}.

Fig.~\ref{fig:chirp CAF} illustrates a CINF structure employed for mitigation of wideband outlier noise affecting a linear chirp signal. Let us first note that the intermittently nonlinear filter used in this example is a feedback-based Analog Differential Clipper (ADiC) described in~\cite{Nikitin19hidden, Nikitin18ADiC-ICC} and in~Section~\ref{subsec:ADiC}, and thus we shall call this particular CINF a {\em Complementary ADiC Filter\/} (CAF).

In Fig.~\ref{fig:chirp CAF}, the output~I of the wideband front-end filter consists of the chirp signal of interest~$x(t)$ and the wideband noise with non-outlier and outlier components~$n(t)$ and~$i(t)$, respectively. Since the stopband of the bandstop filter is relatively narrow in comparison with the passband of the front-end filter, the outlier structure of the noise will be mainly preserved in the output~III of the bandstop filter. However, the bandstop filter significantly reduces the average slew rate of the chirp signal, making the outlier component~$i(t)$ more distinguishable and facilitating its efficient mitigation by an ADiC. As the result, the CAF output~V (that is the sum of the ADiC output~IV and the output~II of the bandpass filter) will consist of the effectively unmodified signal~$x(t)$ and the wideband noise with a reduced outlier component, $n(t)\!+\!\delta{i}(t)$, both delayed by the group delay~$\Delta{t}$ of the bandpass filter. In Fig.~\ref{fig:chirp CAF}, the traces marked by ``$\Delta$" show the respective differences between the filtered signal+noise mixtures and the delayed input signal~$x(t\!-\!\Delta{t})$ (without noise).

Further, Fig.~\ref{fig:chirp example} compares the outcomes of a bandpass filter applied to a chirp signal affected by outlier noise, with and without a CAF deployed ahead of the bandpass filter. The traces marked by ``$\Delta$" show the respective noises in the passband, illustrating the effectiveness of CAF-based outlier interference mitigation.

\begin{figure}[!t]
\centering{\includegraphics[width=8.6cm]{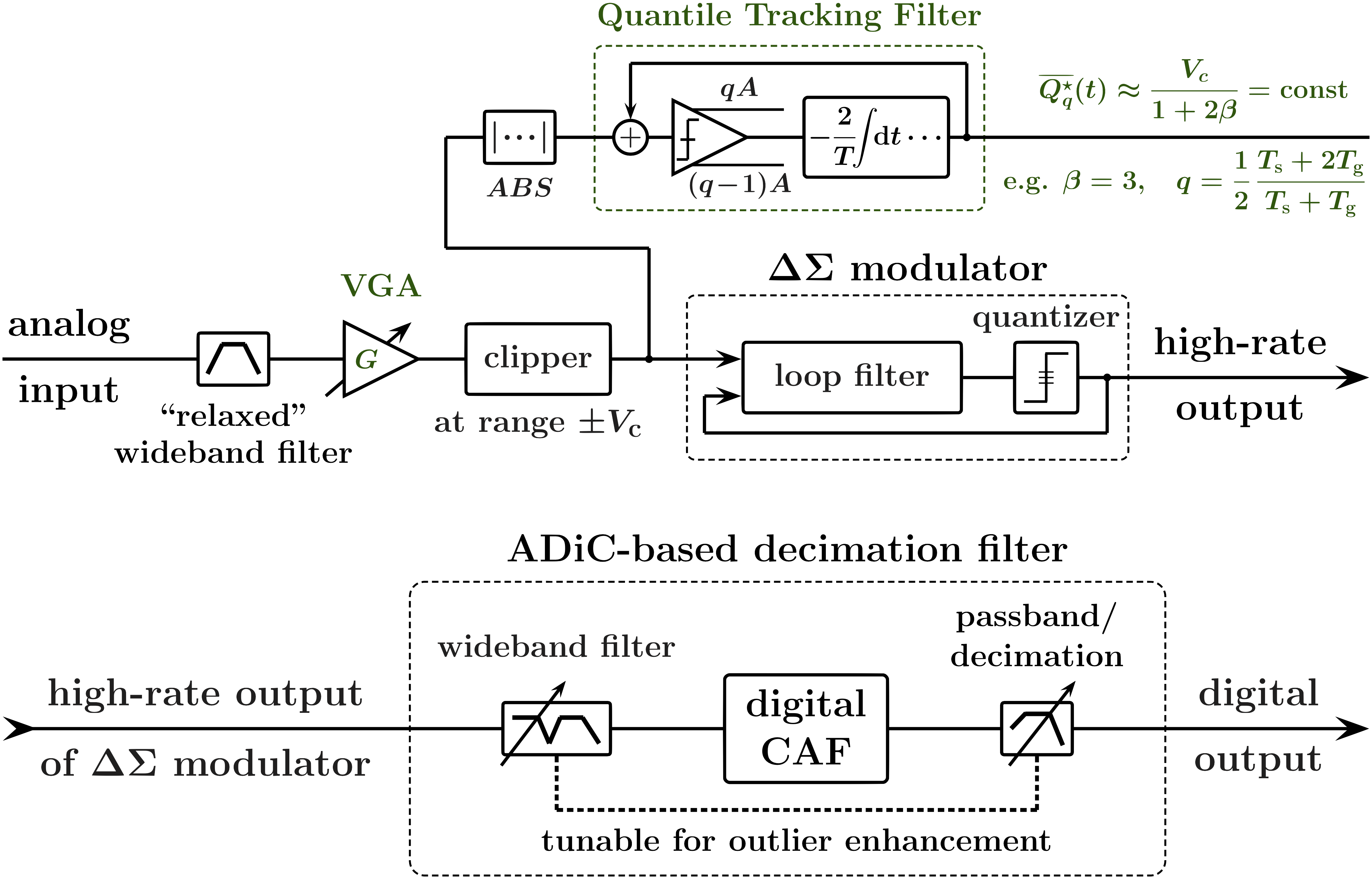}}
\caption{{\bf Example of practical CAF deployment for mitigation of outlier interference during analog-to-digital conversion.~}%
In subsequent simulations, all QTF, CAF, and linear filter parameters are constant.
\label{fig:OFDM setup}}
\end{figure}

\section{Practical CAF implementation and example of its use in OFDM-based systems} \label{sec:practical CAF}
Fig.~\ref{fig:OFDM setup} provides a block diagram of a practical CAF-based system for mitigation of outlier interference in the process of analog-to-digital conversion (ADC).

As discussed in~\cite{Nikitin19hidden} and in Section~\ref{sec:methodology}, efficient mitigation of wideband outlier noise requires availability of a sufficiently broad excess band, and thus the respectively high ADC sampling rate. In addition, the concept of ADiC-based filtering relies on continuous-time (analog) operations such as differentiation, antidifferentiation, and analog convolution. Hence the sampling rate needs to be further increased so that analog differentiation can be replaced by its accurate finite-difference approximation, to enable ``effectively analog" processing. In the implementation shown in Fig.~\ref{fig:OFDM setup}, we use inherently high oversampling rate of a $\Delta\Sigma$~ADC to trade amplitude resolution for higher sampling rate and thus enable efficient digital ADiC-based filtering. Commonly, $\Delta\Sigma$ ADCs are used for converting analog signals over a wide range of frequencies, from DC to several megahertz. These converters comprise a highly oversampling modulator followed by a digital/decimation filter that together produce a high-resolution digital output~\cite{Bourdopoulos03Delta-Sigma, DataConversionHandbook, Geerts06design}.

The high sampling rate allows the use of ``relaxed," wideband antialiasing filtering to ensure the availability of sufficiently wide excess band. As a practical matter, a wideband filter with a flat group delay and a small time-bandwidth product (e.g. with a Bessel response) should be used in order to increase the mitigable rates. Further, a simple analog clipper should be employed ahead of the $\Delta\Sigma$~modulator to limit the magnitude of excessively strong outliers in the input signal to the range~$\pm V_{\rm c}$ that is smaller than that of the quantizer. Such a clipper will prevent the modulator from saturation. In Fig.~\ref{fig:OFDM setup}, a {\em robust\/} automatic gain control circuit adjusts the gain~$G$ of the variable-gain amplifier (VGA) to maintain a constant output of a properly configured Quantile Tracking Filter circuit (see Section~\ref{subsec:QTFs}) applied to the absolute value of the clipper output. This ensures that only large noise outliers are clipped, and not the outliers of the signal itself, e.g. outliers in high-crest-factor signals such as OFDM.

The low (e.g. 1-bit) amplitude resolution of the output of the $\Delta\Sigma$ modulator does not allow direct application of a digital ADiC. However, since the oversampling rate is significantly higher (e.g. by two to three orders of magnitude) than the Nyquist rate of the signal of interest, a wideband digital filter can be first applied to the output of the quantizer to enable the ADiC-based outlier filtering. To reduce computations and memory requirements, such a filter can be an IIR filter. For instance, for a 1-bit $\Delta\Sigma$ modulator with a 20\,MHz clock, and a required 100\,kS/s decimated output, the bandwidth of the wideband IIR filter ahead of the CAF in Fig.~\ref{fig:OFDM setup} can be about 500\,kHz. Furthermore, the analog antialiasing filter and the wideband IIR filter should be co-designed to ensure the desired excess band response in both time and frequency domains. For example, if both the analog antialiasing and the wideband IIR digital filters are 2nd~order lowpass filters, their corner frequencies and the quality factors can be chosen to ensure that the combined response of these cascaded filters is that of the 4th~order Bessel-Thomson filter~\cite{Schaumann01DesignOfAnalogFilters, Proakis06digital}.

\subsection{Quantile Tracking Filters (QTFs) for establishing robust front-end clipping levels~$\pm V_{\rm c}$ and ADiC fences $\alpha_+$ and $\alpha_-$} \label{subsec:QTFs}
A robust time-varying range ${\left[\alpha_-(t),\alpha_+(t)\right]}$ that excludes outliers of a signal can be obtained as a range between {\em Tukey's fences\/}~\cite{Tukey77exploratory} constructed as linear combinations of the 1st ($Q_{[1]}$) and the 3rd ($Q_{[3]}$) quartiles of the signal in a moving time window:
\beginlabel{equation}{eq:Tukey's range}
  [\alpha_-,\alpha_+] \!=\! {\big [}Q_{[1]}\!-\!\beta\left(Q_{[3]}\!-\!Q_{[1]}\right)\!,Q_{[3]}\!+\!\beta\left(Q_{[3]}\!-\!Q_{[1]}\right)\!{\big ]},
\end{equation}
where $\alpha_+$, $\alpha_-$, $Q_{[1]}$, and $Q_{[3]}$ are time-varying quantities, and $\beta$ is a scaling parameter of order unity (e.g. $\beta=1.5$). In practical analog and/or real-time digital implementations, approximations for the time-varying quartile values can be obtained by means of Quantile Tracking Filters (QTFs) described in detail in~\cite{Nikitin17nonlinear, Nikitin19ADiCpatentCIPs, Nikitin18ADiC-ICC}. In brief, the signal~$Q_q(t)$ that is related to a given input~$y(t)$ by the equation
\beginlabel{equation}{eq:Qq}
  \frac{\d}{\d{t}}\, Q_q = \frac{A}{T}\, \left[\sgn(y\!-\!Q_q) + 2q-1\right],
\end{equation}
where~$A$ is a parameter with the same units as~$y$ and $Q_q$, and~$T$ is a constant with the units of time, can be used to approximate (``track") the $q$-th~quantile of $y(t)$ for the purpose of establishing a robust range ${[\alpha_-,\alpha_+]}$. (See~\cite{Nikitin03signal, Nikitin04adaptive} for discussion of quantiles of continuous signals.) From equation~(\ref{eq:Tukey's range}), for a signal with symmetrical amplitude distribution the range that excludes outliers can also be obtained as ${[\alpha_-,\alpha_+] = [-\alpha,\alpha]}$, where~$\alpha$ is given by
\beginlabel{equation}{eq:Tukey's range sym}
  \alpha = (1+2\beta) Q^\star_{[2]}\,,
\end{equation}
and where~$Q^\star_{[2]}=Q^\star_{1/2}$ is the 2nd~quartile (median, $q=1/2$) of the absolute value of the signal.

\begin{figure}[!t]
\centering{\includegraphics[width=8.6cm]{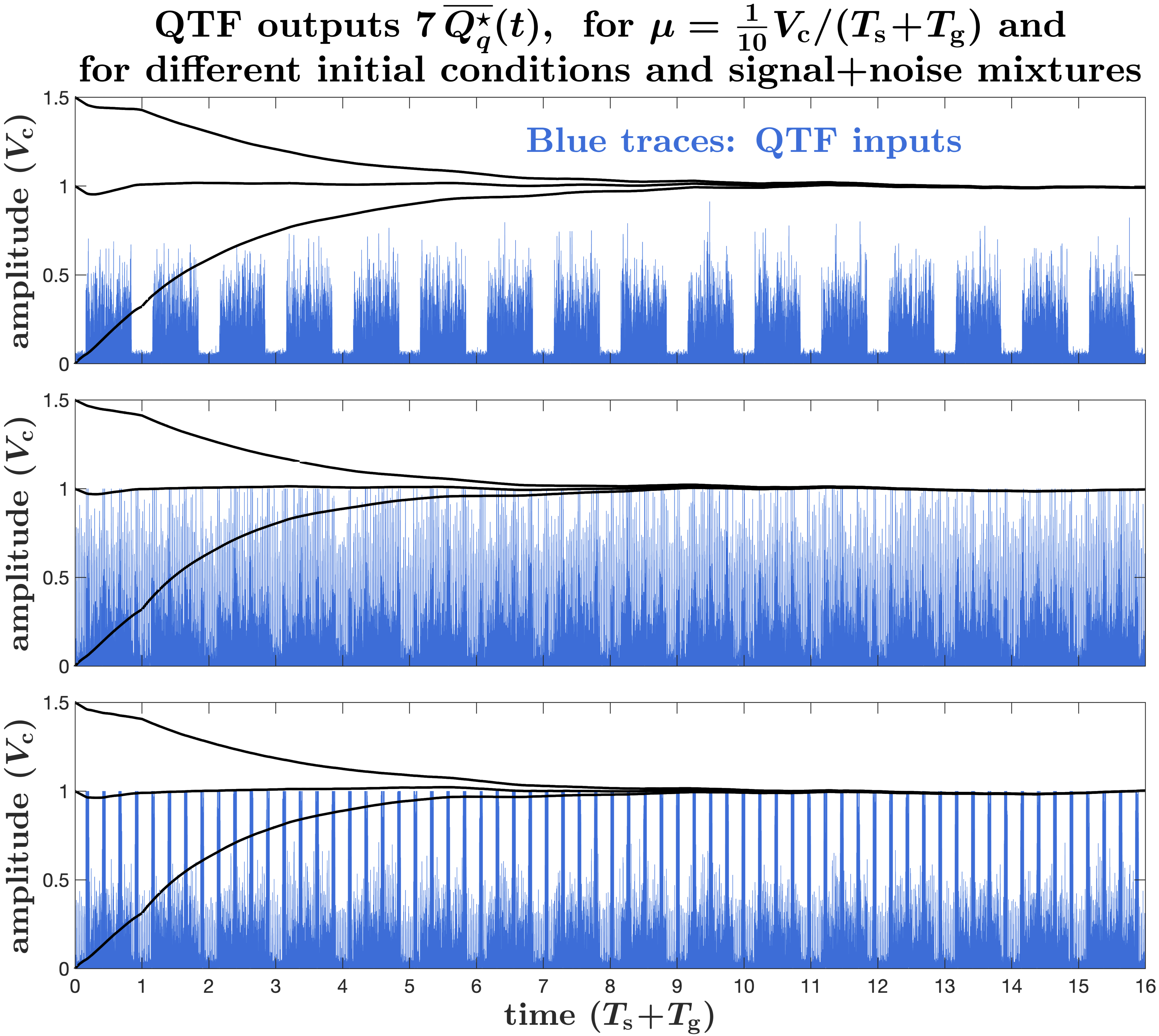}}
\caption{{\bf Examples of QTF outputs convergence to steady states for different initial conditions and signal+noise mixtures.~}%
Averaging of $Q^\star_q(t)$ is performed in moving time window of width~$T_{\rm s}\!+\!T_{\rm g}$.
\label{fig:QTF outputs}}
\end{figure}

Due to their high crest factor, for OFDM signals the scaling parameter~$\beta$ in~(\ref{eq:Tukey's range sym}) should be relatively large (e.g. $\beta=3$) to ensure that the outliers in the signal itself are not clipped. In addition, since the OFDM symbols of duration~$T_{\rm s}$ are separated by the guard intervals of duration~$T_{\rm g}$, the quantile value $q=1/2$ in~(\ref{eq:Tukey's range sym}) should be increased to ${q=\half(T_{\rm s}+2T_{\rm g})/(T_{\rm s}+T_{\rm g})}$ to account for the guard intervals. Further, for a balanced compromise between robustness to outliers and the response speed, the QTF slew rate parameter~$\mu=A/T$ can be chosen to be relatively large, e.g. ${\mu=\frac{1}{10}V_{\rm c}/(T_{\rm s}+T_{\rm g})}$, and the average QTF output~$\overline{Q^\star_q}(t)$ can then be obtained in a moving time window of width~$(T_{\rm s}\!+\!T_{\rm g})n$, where~$n$ is an integer. Fig.\,\ref{fig:QTF outputs} illustrates QTF outputs convergence to steady states for different initial conditions (different black lines) and signal+noise mixtures, for ${\mu=\frac{1}{10}V_{\rm c}/(T_{\rm s}+T_{\rm g})}$ and $Q^\star_q(t)$ averaging in a moving time window of width~$T_{\rm s}\!+\!T_{\rm g}$. In the upper panel, the QTF input corresponds to an OFDM signal affected by the thermal noise only, with 30\,dB SNR in the OFDM passband, and the signal is not being clipped. In the middle and the lower panels, outlier interference with different compositions and intensities is added to the OFDM signal, and the excessively strong outliers are limited to the range~$\pm V_{\rm c}$.

\subsection{Feedback-based ADiC} \label{subsec:ADiC}
The basic concept of an ADiC as a particular type of an intermittently nonlinear filter can be briefly described as follows~\cite{Nikitin19hidden}:
First, we establish a robust range that excludes noise outliers while including the signal of interest; then, we replace the values that extend outside of the range with those in mid-range. When we are not constrained by the needs for either analog or wideband, high-rate real-time digital processing, in the digital domain these steps can perhaps be accomplished by a {\it Hampel filter\/}~\cite{Hampel74influence} or by one of its variants~\cite{Pearson16generalized}. In a Hampel filter the ``mid-range" is calculated as a windowed median of the input, and the range is determined as a scaled absolute deviation about this windowed median. However, Hampel filtering cannot be performed in the analog domain, and/or it becomes prohibitively expensive in high-rate real-time digital processing. Unlike Hampel filtering, ADiCs rely on continuous-time operations that allow wideband analog and/or high-rate digital implementations in real time.

\begin{figure}[!b]
\centering{\includegraphics[width=8.6cm]{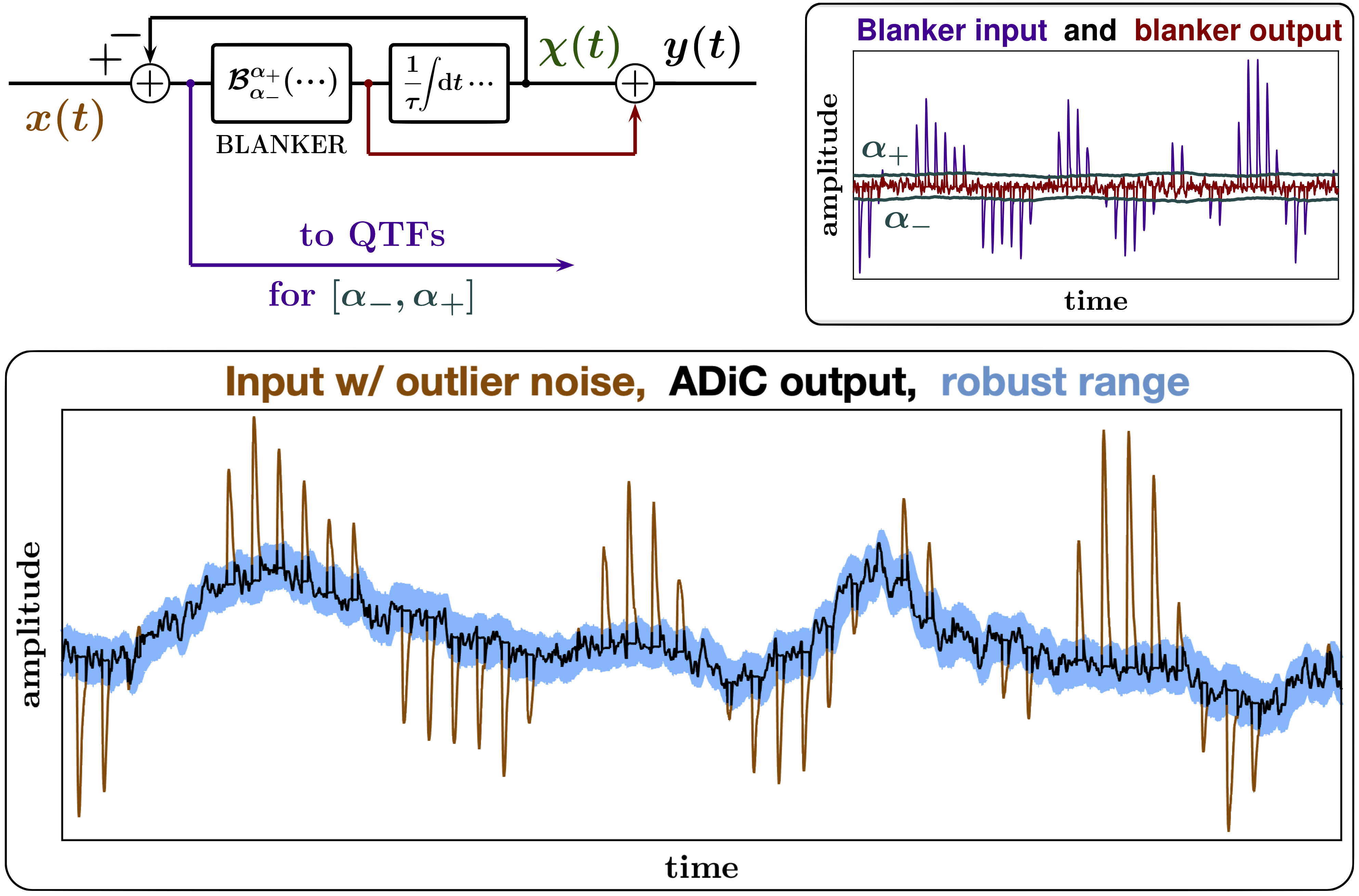}}
\caption{{\bf\boldmath Feedback-based ADiC replacing outliers with~$\chi(t)$.~}
Reproduced from~\cite{Nikitin19hidden}.
\label{fig:ADiC}}
\end{figure}

Fig.\,\ref{fig:ADiC} presents a feedback-based ADiC variant that has a number of practical advantages and is well suited for mitigation of hidden outlier noise~\cite{Nikitin19hidden}. As the diagram in the upper left of the figure shows, the ADiC output~$y(t)$ can be described as
\beginlabel{equation}{eq:ADiC equation}
  \left\{
  \begin{array}{cc}
    \!\! y(t) = \chi(t) + \tau\dot{\chi}(t)\\[1mm]
    \!\! \displaystyle \dot{\chi}(t) = \frac{1}{\tau}\, \BalphaPM\left(x(t)\!-\!\chi(t) \right)
  \end{array}\right.,
\end{equation}
where~$x(t)$ is the input signal, $\chi(t)$~is the {\em differential clipping level\/} (DCL), the {\em blanking function\/}~$\BalphaPM(x)$ is a particular type of an {\em influence function}~\cite{Hampel74influence} that is defined as
\beginlabel{equation}{eq:clipping function}
  \BalphaPM(x)  = \left\{
  \begin{array}{cc}
    \!\! x & \mbox{for} \quad \alpha_- \le x \le \alpha_+\\
    \!\! 0 & \mbox{otherwise}
  \end{array}\right.,
\end{equation}
and where~$[\alpha_-,\alpha_+]$ is a robust range for the {\em difference signal\/}~${x(t)-\chi(t)}$ (the {\em blanking range\/}). Thus such an ADiC is an intermittently nonlinear filter that outputs the DCL~$\chi(t)$ only when outliers in the difference signal are detected, performing outlier noise mitigation without modifying the input signal otherwise. For the range fences such that ${\alpha_- \le x(t)\!-\!\chi(t) \le \alpha_+}$ for all~$t$, the DCL~$\chi(t)$ is the output of a 1st~order linear lowpass filter with the 3\,dB corner frequency~$1/(2\pi\tau)$. However, when an outlier of the difference signal is encountered, the rate of change of~${\chi(t)}$ is zero and the DCL maintains its previous value for the duration of the outlier.

\subsection{CAF vs. linear: ``No harm" condition and effect on SNRs and channel capacities} \label{subsec:capacities}
As discussed in Section~\ref{sec:methodology}, apparent outliers in the interference can disappear and reappear due to various filtering effects, including fading and multipass, as the signal propagates through media and/or the signal processing chain. Although various combinations of linear filters can be used to increase the difference between the temporal and/or amplitude structures of the interference and the signal of interest, enhancing the outlier components of the interference and facilitating its mitigation by intermittently nonlinear filtering, such an approach may not be easily accomplished in practice. For example, it may require a sufficient {\em a priori\/} knowledge of the interference structure (e.g. the presence of adjacent channel interference), and/or employing various machine learning and optimization-based approaches. The ``interference enhancement" may be especially difficult to accomplish in complex, highly nonstationary interference scenarios, e.g. in mobile and cognitive communication systems where the transmitter positions, powers, signal waveforms, and/or spectrum allocations vary dynamically. Thus one of the main requirements for the CINF-based interference mitigation is its ability to operate under the ``no harm" constraint such that nonlinear filtering does not degrade the resulting signal quality, as compared with the linear filtering, for any signal+noise mixtures. That is, while providing resistance to outlier noise, in the absence of such noise CINFs must behave effectively linearly, avoiding the detrimental effects, such as distortions and instabilities, often associated with nonlinear filtering.

Although it is perhaps unrealistic to require that any ``default" set of CINF parameters satisfies the ``no harm" constraint while improving the signal quality for all conceivable interference conditions, this constraint can always be met, for any particular interference scenario, in an ADiC-based filtering. Indeed, in the limit of a wide ADiC's blanking range such that ${\alpha_- \le x(t)\!-\!\chi(t) \le \alpha_+}$ for all~$t$, a CAF becomes effectively an allpass filter with a constant group delay, and it will not degrade the resulting signal quality, as compared with the linear filtering, for any signal+noise mixtures. Note, however, that when a CAF does improve the signal quality, its performance can be further enhanced by optimizing its parameters.

\begin{figure*}[!t]
\centering{\includegraphics[width=17.7cm]{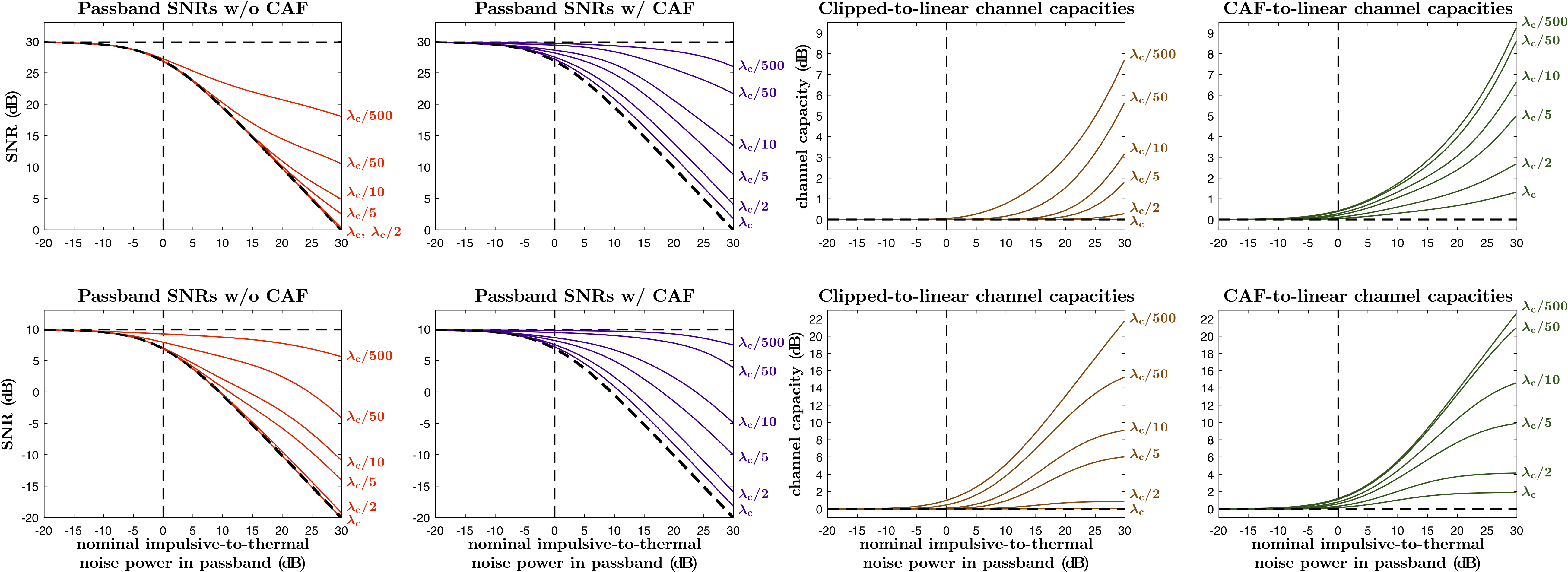}}
\caption{{\bf\boldmath Poisson noise with normally distributed amplitudes:~}
CAF-based filtering following analog clipper noticeably increases effectiveness of mitigation, especially for high SNRs and event occurrence rates.
\label{fig:OFDM Poisson}}
\end{figure*}
\begin{figure*}[!t]
\centering{\includegraphics[width=17.7cm]{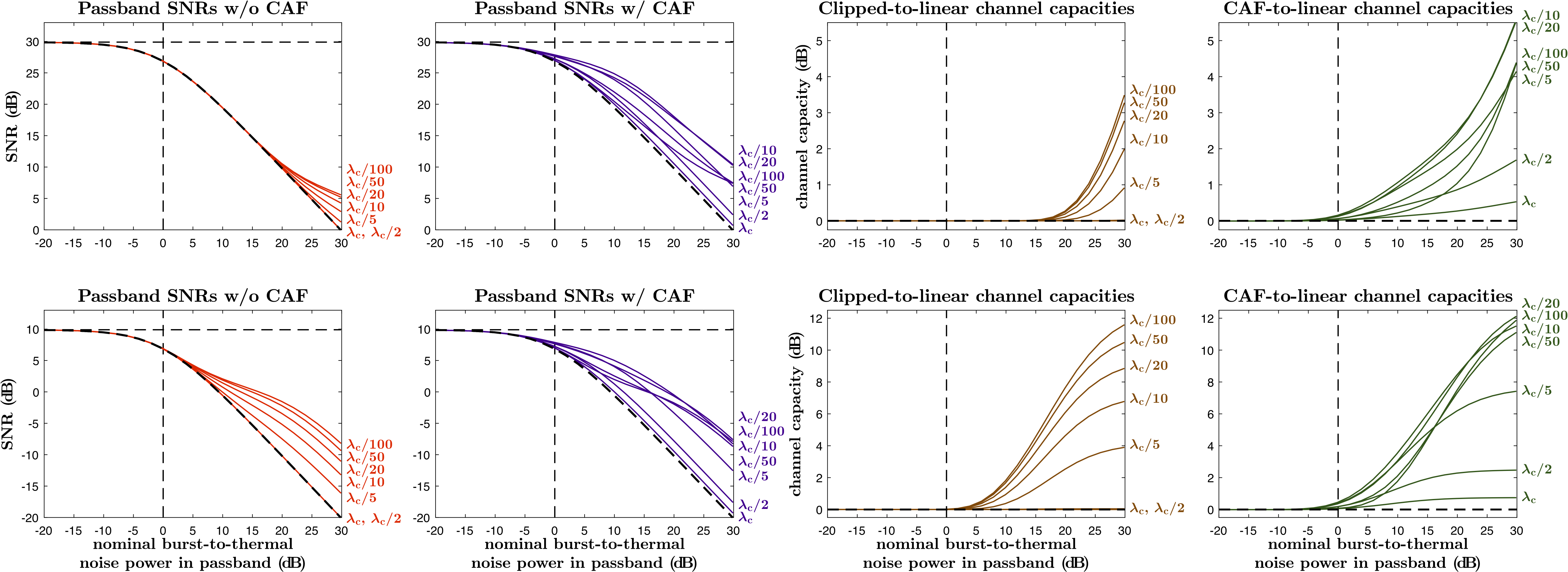}}
\caption{{\bf\boldmath Periodic Gaussian bursts with 10\% duty cycle:~}
CAF-based filtering following analog clipper significantly further improves signal quality and extends mitigability, but its effectiveness is no longer monotonic with respect to outlier occurrence rates (since burst duration is inversely proportional to rate).
\label{fig:OFDM bursts}}
\end{figure*}
\begin{figure*}[!t]
\centering{\includegraphics[width=17.7cm]{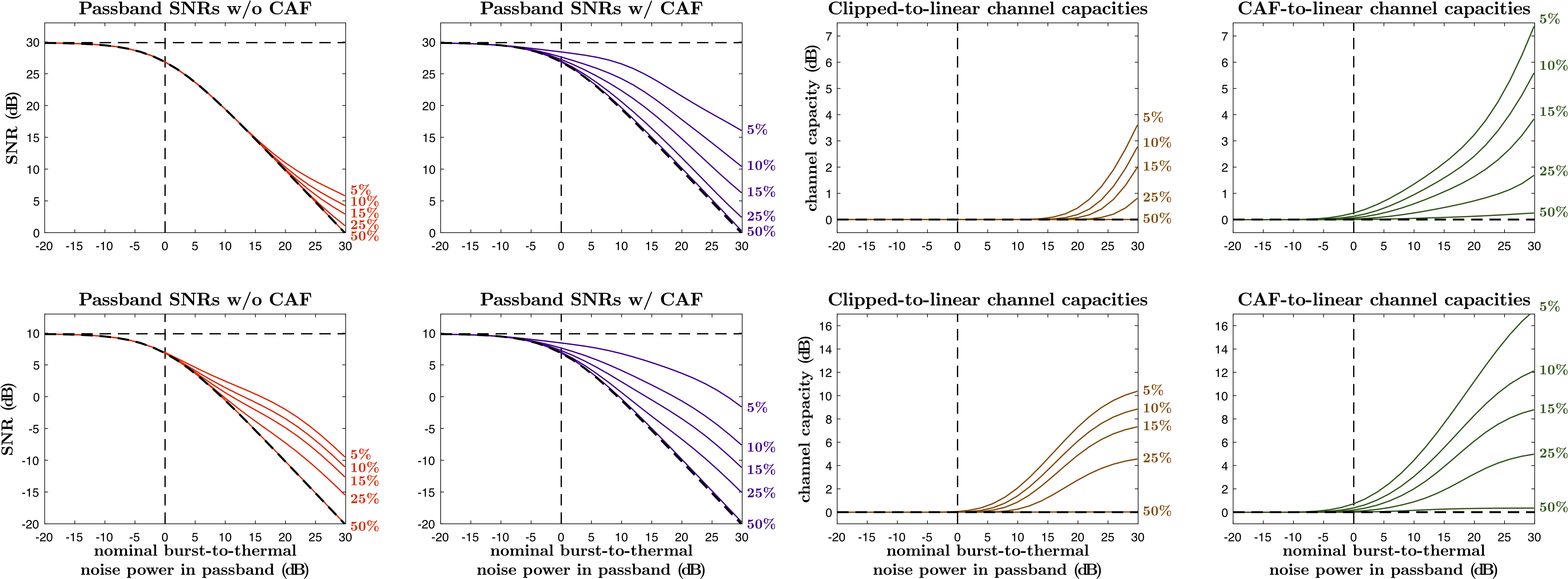}}
\caption{{\bf\boldmath Periodic Gaussian bursts with $\lambda\!=\!\lambda_{\rm c}/20$ and different duty cycles:~}
For bursts with duty cycles larger than~50\% CAF-based filtering with default parameters becomes ineffective.
\label{fig:OFDM bursts DC}}
\end{figure*}
\begin{figure}[!t]
\centering{\includegraphics[width=8.6cm]{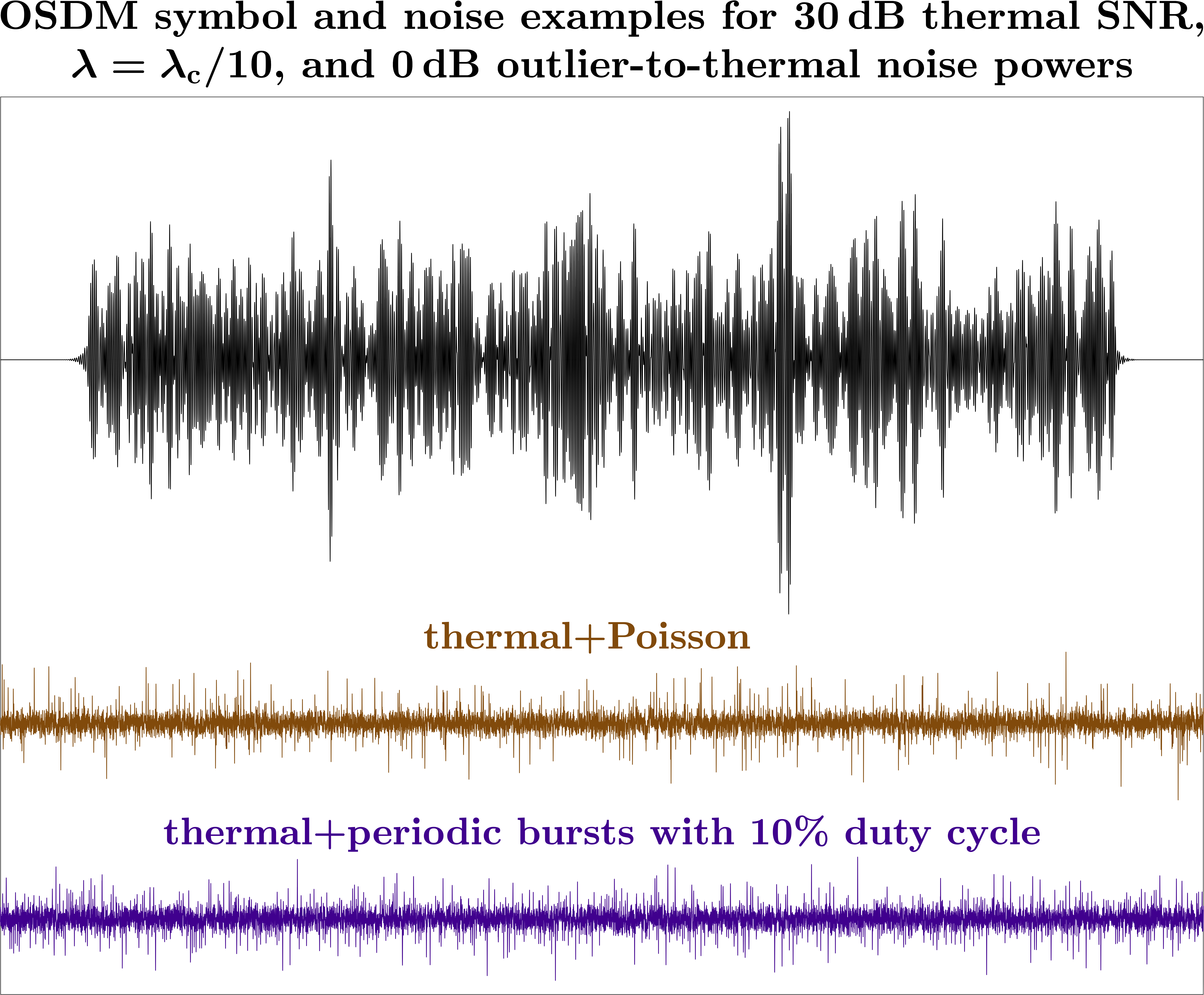}}
\caption{Example of OFDM signal and noise traces after wideband front-end filter, at onset of outlier interference mitigability.
\label{fig:noise examples}}
\end{figure}

Figs.~\ref{fig:OFDM Poisson} through~\ref{fig:OFDM bursts DC} illustrate and quantify the effectiveness of the outlier interference mitigation provided by the CAF-based filtering arrangement outlined in Fig.~\ref{fig:OFDM setup}. The signal of interest models a simplified OFDM-based signal and consists of QPSK-modulated symbols of duration~$T_{\rm s}$ separated by guard intervals of duration~${T_{\rm g}\approx 0.46T_{\rm s}}$. The OFDM central frequency is~$f_{\rm c}$, the bandwidth is~$f_{\rm c}/3$, and the number of subcarriers is~$256$. Various mixtures of wideband thermal and outlier noise are added to the signal, with the outlier interference such as Poisson noise with normally distributed amplitudes (Fig.~\ref{fig:OFDM Poisson}), and periodic Gaussian bursts with different rates and duty cycles (Figs.~\ref{fig:OFDM bursts} and~\ref{fig:OFDM bursts DC}). In the simulations, two base values of the thermal noise SNRs in the OFDM passband were used, 30\,dB and 10\,dB, and the power of the added outlier interference (measured in relation to the thermal noise power) varies in a 50\,dB range, from $-20\,$dB to 30\,dB.

The front-end wideband filter is a flat-group-delay (Bessel) filter with the nominal 3\,dB passband~$[f_{\rm c}/75,4f_{\rm c}]$, and its time-bandwidth product is approximately that of a Gaussian filter, ${2\log_2(2)/\pi}$. Thus~$\lambda_{\rm c}\approx 9 f_{\rm c}$ is the ``pileup threshold" rate of the front-end filter~\cite{Nikitin19hidden}. As discussed in~\cite{Nikitin19hidden}, for outlier arrival rates significantly above~$\lambda_{\rm c}$ the outliers in the interference effectively disappear due to the pileup effect, and can no longer be mitigated.

As shown in Fig.~\ref{fig:OFDM setup}, an analog clipper, along with a QTF-based gain control circuit, is employed ahead of the $\Delta\Sigma$ modulator to limit the magnitude of excessively strong outliers to the range~$\pm V_{\rm c}$. As discussed earlier in this section, the QTF-based range control ensures that only large noise outliers are clipped, and not the outliers of the signal itself. This by itself provides resilience to high-power, low rate outlier interference, improving the resulting signal quality in the presence of such interference. Hence the comparison of the effects of the front-end clipping alone, without the CAF-based filtering, is included in the simulation results alongside the effects of the CAF-based filtering.
Since, for the noise compositions used in the simulations, there is no need for the ``interference enhancement" before the CAF, the wideband filter in the ADiC-based decimation filter of Fig.~\ref{fig:OFDM setup} is a simple 2nd~order lowpass Bessel filter, and there are no modifications in the decimation filter.

Figs.~\ref{fig:OFDM Poisson}, \ref{fig:OFDM bursts}~and~\ref{fig:OFDM bursts DC} show the comparative improvements in the passband SNRs and in the channel capacities, as functions of the outlier-to-thermal noise power in the OFDM passband, for different outlier noise compositions and moderate (10\,dB) and high (30\,dB) thermal noise SNRs. Since CAF-based filtering removes noise outliers, the passband noise after such filtering is effectively Gaussian, and the Shannon formula~\cite{Shannon49communication} can be used to calculate the limit on the channel capacity. However, the passband noise without the CAF in the signal chain may not be Gaussian, especially for low outlier rates and high outlier interference powers. Nevertheless, we still use the Shannon formula as a proxy measure for the capacity of the linear channel, to quantify the comparative signal quality improvement.
In all simulations, the QTF, CAF, and linear filter parameters are constant, and no parameter optimization is performed.

Fig.~\ref{fig:noise examples} illustrates the comparative time domain appearances of the OFDM symbol and the noise traces, after the wideband front-end filter, at the approximate onset of the CAF-based mitigability of outlier interference. A further increase in the power of the outlier component will result in significantly larger relative improvement in the signal quality.

\section{Conclusion} \label{sec:conclusion}
In this paper, we provide a brief overview of the methodology and tools for real-time mitigation of outlier interference in general and ``hidden" wideband outlier noise in particular. Either used by itself, or in combination with subsequent mitigation techniques, this approach provides interference reduction levels otherwise unattainable, with the effects, depending on particular interference scenarios, ranging from ``no harm" to considerable. While the main focus of this filtering technique is to serve as a ``first line of defense" against wideband interference ahead of, or in the process of, analog-to-digital conversion, it can also be used, given some {\em a priori} knowledge of the signal of interest's structure, to reduce outlier interference that is confined to the signal's band.

In addition to addressing ``hidden interference" scenarios, the distinct feature of the proposed approach is that it capitalizes on the ``excess band" observation of interference for its efficient in-band mitigation by intermittently nonlinear filters. This significantly extends the mitigation range, in terms of both the rates of the outlier events and the mitigable SNRs, in comparison with the mitigation techniques focused on the apparent in-band effects of outlier interference.

The CINF-based structure described in the paper is mostly ``blind" as it does not rely on any assumptions for the underlying interference beyond its ``inherent" outlier structure, and it is adaptable to nonstationary signal and noise conditions and to various complex signal and interference mixtures. Thus it can be successfully used to suppress interference from diverse sources, including the RF co-site interference and the platform noise generated by on-board digital circuits, clocks, buses, and switching power supplies. It can also help to address multiple spectrum sharing and coexistence applications (e.g. radar-communications, radar-radar, narrowband/UWB, etc.), including those in dual function systems (e.g. when using radar and communications as mutual signals of opportunity). This filtering paradigm can further benefit various other military, scientific, industrial, and consumer systems such as sensor/sensor networks and coherent imaging systems, sonar and underwater acoustic communications, auditory tactical communications, radiation detection, powerline communications, navigation and time-of-arrival techniques, and many others. Finally, various embodiments of the presented filtering structure can be integrated into, and manufactured as IC components for use in different products, e.g. as A/D converters with incorporated interference suppression.

\section*{Acknowledgment}
The authors would like to thank
Keith~W. Cunningham of Atkinson Aeronautics \& Technology Inc., Fredericksburg, VA,
and Kyle~D. Tidball of Mid-Continent Instruments and Avionics, Wichita, KS,
for their valuable suggestions and critical comments.

\small
\input{ADiC_MILCOM19.bbl}

\end{document}